\documentclass[12pt]{article}% scrbook, scrreprt, scrlettr, scrartcl
\usepackage[body={17cm, 23cm, centered}]{geometry}
\usepackage{epsfig,verbatim,cite}
\usepackage{lmodern}
\usepackage{braket}
\usepackage[T1]{fontenc}
\usepackage{mathtext,marvosym,textcomp}
\usepackage{mathtools}
\usepackage{slashed}
\usepackage[usenames,dvipsnames,svgnames,table]{xcolor}
\usepackage{tikz}
\usepackage[normalem]{ulem}
\usepackage{tocloft}
\usepackage{epsfig,amsmath,amsfonts,amssymb,arydshln}
\usepackage[makeroom]{cancel}
\usepackage{multirow}

 \usepackage{relsize}

 \usepackage[debug,pageanchor=false]{hyperref}
\hypersetup{colorlinks=true,linktocpage,breaklinks,
            urlcolor=blue,
            linkcolor=red,
            citecolor=blue
            }

\numberwithin{equation}{section}

\def\ph{\phantom}

\newcommand\bqa {\begin{eqnarray}}
\newcommand\eqa {\end{eqnarray}}

\newcommand{\bear}{\begin{array}}
\newcommand{\enar}{\end{array}}

\newcommand{\be}{\begin{equation}}
\newcommand{\ee}{\end{equation}}
\newcommand{\bea}{\begin{eqnarray}}
\newcommand{\eea}{\end{eqnarray}}

\usepackage{tikz}
\usetikzlibrary{arrows}
\usetikzlibrary{shapes.misc}
\usepackage{tikz-cd}

\usetikzlibrary{arrows,shapes,positioning, fit}
\tikzstyle{every picture}+=[remember picture]
\tikzstyle{na} = [baseline=-.5ex]
\tikzstyle{format} = [rounded rectangle,
                      thick,
                      minimum size=1cm,
                      draw=red!00!black!80,
                      top color=white,
                      bottom color=red!00!black!00,
                      on grid]
\tikzstyle{format1} = [rectangle,
                      thick,
                      minimum size=1cm,
                      draw=red!00!black!80,
                      top color=white,
                      bottom color=red!00!black!00,
                      on grid]
\tikzstyle{format0} = [rounded rectangle,
                      thick,
                      minimum size=1.2cm,
                      draw=blue!00!black!80,
                      top color=white,
                      bottom color=blue!00!black!00,
                      text centered]
\tikzstyle{formatd} = [rounded rectangle,
                      thick,
                      dashed,
                      minimum size=1cm,
                      draw=blue!50!black!80,
                      top color=white,
                      bottom color=blue!00!black!00,
                      text centered]
\tikzstyle{format1d} = [rounded rectangle,
                      thick,
                      dashed,
                      minimum size=2cm,
                      draw=blue!50!black!80,
                      top color=white,
                      bottom color=blue!00!black!00,
                      text centered]

\tikzset{cross/.style={cross out, draw=black, minimum size=2*(#1-\pgflinewidth), inner sep=0pt, outer sep=0pt},
    cross/.default={5pt}}
\usepackage[ normalem]{ulem}
\usepackage{tocloft}

\begin{document}
\renewcommand{\contentsname}{}
\renewcommand{\refname}{\begin{center}References\end{center}}
\renewcommand{\abstractname}{\begin{center}\footnotesize{\bf Abstract}\end{center}}

\begin{titlepage}
\ph{preprint}

\vfill

\begin{center}
   \baselineskip=16pt
   {\large \bf More on thermal holographic RG flows
   in a 3D gauged supergravity
   }
   \vskip 2cm
    Anastasia A. Golubtsova$^{a,b}$\footnote{\tt golubtsova@theor.jinr.ru },
      Eric Gourgoulhon$^{c,d}$\footnote{\tt eric.gourgoulhon@obspm.fr }
     and
    Mikhail A. Podoinitsyn$^a$\footnote{\tt mpod@theor.jinr.ru}
       \vskip .6cm
             \begin{small}
                \vskip .6cm
             \begin{small}
                          {\it
                          $^a$  Bogoliubov Laboratory of Theoretical Physics, JINR,\\
Joliot-Curie str. 6,  Dubna, 141980  Russia  \\
                          $^b$ Steklov Mathematical Institute, Russian Academy of Sciences\\
Gubkina str. 8, 119991 Moscow, Russia\\
                          $^c$ LUTH, UMR CNRS 8102, Observatoire de Paris, Universit\'e PSL, Universit\'e Paris Cit\'e, 5 place Jules Janssen, 92190 Meudon, France\\
                          $^d$ Laboratoire de Math\'ematiques de Bretagne Atlantique, UMR CNRS 6205, Universit\'e de Bretagne Occidentale, 6 avenue Victor Le Gorgeu, 29200 Brest, France}\\

\end{small}
\end{small}
\end{center}
\vfill
\begin{center}
\textbf{Abstract}
\end{center}
\begin{quote}
We continue our studies of holographic renormalization group (RG) flows for a 3d truncated supergravity model, the scalar potential of which can have either one or three extrema depending on  the radius of the target manifold. We construct numerically and analytically thermal holographic RG flows, which are described by asymptotically AdS$_3$ black holes (non-rotating BTZ) characterized by the value of the scalar field on the horizon. We find two classes of RG flows with monotonic and non-monotonic behavior of the scalar field. The first one exists for both types of the potential, while the second one appears only for the potential with three extrema. For the slowly changing scalar field  we find a special class of RG flows,  which are described by the scalar field in the BTZ black hole geometry. For such flows we present an analytic solution for the scalar field from the horizon to the boundary. We discuss thermodynamical properties of the constructed holographic RG flows.
\end{quote}

\vfill
\setcounter{footnote}{0}
\end{titlepage}

\tableofcontents

\setcounter{page}{2}
\newpage
\section{Introduction}

The holographic duality \cite{Maldacena:1997re,Witten:1998qj,Gubser:1998bc} is a tool that allows to describe various aspects of physical systems under strong coupling using gravity.

In the holographic prescription a RG flow \cite{Akhmedov:1998vf} can be represented as a Poincar\'e invariant (domain wall) solution with AdS asymptotics and appropriate boundary conditions for the field content \cite{deBoer:1999tgo,Freedman:1999gp,deHaro:2000vlm,Bianchi:2001kw,Bianchi:2001de,Skenderis:2002wp,Papadimitriou:2004ap,Papadimitriou:2004rz,Papadimitriou:2007sj}. The  extrema of the scalar potential, where the geometries are AdS spacetimes,  associate with fixed points of the dual field theory, which are deformed either by a relevant operator or by non-zero VEV of the operator. %The case when the flow runs to infinity, i.e. \phi points of the potential can be also implemented into the holographic picture and can be related with fixed points with scaling.

Thermal RG flows encodes an information on a physical system and may shed light on puzzling properties of exotic RG flows. Finite temperature generalization of holographic fixed points is represented by AdS black holes. %Thus, the Hawking temperature corresponds to a certain.
Holographic RG flows were studied broadly  for the phenomenological applications  in \cite{Gursoy:2008za,Donos:2017sba,Arefeva:2018hyo,Arefeva:2018jyu,Gursoy:2018umf,Bea:2018whf,Arefeva:2019qen,Arefeva:2024vom,Faedo:2024zib}.
Another direction of using thermal RG flows in holographic theories is to probe the interior of the black hole in the gravitational dual \cite{Frenkel:2020ysx,Caceres:2022smh,Caceres:2023zft}.

In this paper we continue studies of holographic RG flows at finite temperature in 3d gauged supergravity coupled to a single scalar
field with a potential
\cite{Deger:1999st,Deger:2002hv}, which were constructed in \cite{Golubtsova:2024dad}. Despite a simple form of the model it includes interesting features that can be realized in more sophisticated supergravity constructions. The scalar potential can have  either one or three extrema depending on the value of the parameter related with the radius of the scalar target manifold.   An exact domain wall solution for this model,  describing half-supersymmetric RG flow was found in \cite{Deger:2002hv}. In works \cite{Deger:2004mw, Deger:2006uc}  string solutions for this model were  constructed.
In \cite{ Golubtsova:2022hfk,Arkhipova:2024iem} holographic RG flows  at zero temperature for this model were studied reducing the equations motion to an autonomous dynamical system using the scalar variables \cite{Gursoy:2018umf}. In \cite{Arkhipova:2024iem} for RG flows driven by the relevant operator as well as non-zero VEV  with various boundary conditions (Dirichlet, Neumann, mixed) "pseudo"-superpotentials were found.

Note, that the reconstruction of the (super)gravity solutions from  the analysis of fixed points of the dynamical systems was discussed in \cite{Sonner:2005sj,Arefeva:2018jyu,Golubtsova:2018dfh}. In particular, finding domain wall solutions with zero and finite temperatures for the  truncated supergravity model  were performed in \cite{Golubtsova:2022hfk,Arkhipova:2024iem,Golubtsova:2024dad}. %Stability analysis  of black hole solutions using   global phase portraits was considered in works \cite{Ganguly:2014qia, Cruz:2017ecg}.

Holographic RG flows, constructed in \cite{Golubtsova:2024dad}, are thermal generalization to those from \cite{Golubtsova:2022hfk,Arkhipova:2024iem}.
These flows involve some numerical input.  The equations of motion  are represented as an autonomous  dynamical system  which configuration space is a unit cylinder. In such configuration space near-horizon regions of the solutions are mapped on the line of the boundary of the cylinder. Numerical black hole solutions  are constructed such that for the initial conditions we used the constraints for the coordinates and the value of the scalar field on the horizon. In \cite{Golubtsova:2024dad} it was shown that a separatrix for the constructed asymptotically AdS black holes for the case of the potential with one extremum is the supersymmetric zero-temperature RG flow \cite{Deger:2002hv} between AdS and scaling fixed points, while for the case of the potential with three extrema the separatrix is the non-supersymmetric  zero-temperature RG flow between two AdS fixed points.   Moreover, using the cylinder for the configuration space allows to construct analytic near-horizon solutions, which have AdS asymptotics. From the analysis of the scalar field dynamics it was established special class of the flows, for which one can assume an additional constraint yielding an analytic solution for the scalar field from the boundary to the horizon. An interesting point is that under this constraint the geometry does not have a deformation  and is described by the BTZ metric, while the scalar field is given  by the solution to the hypergeometric equation.

%\textcolor{blue}{Bulk-boundary propagators in the BTZ black  hole and their relations  with the pure AdS$_{3}$ case were discussed in \cite{Keski-Vakkuri:1998gmz}.  The scalar field in AdS$_3$ and bulk-boundary  propagators \cite{Freedman:1998tz,Balasubramanian:1998sn}.}

The outline of the paper is as follows.
In Section \ref{sec:setup} we review the holographic model, which scalar potential depends on the parameter $a^2$, and write down equations of motion using a black hole ansatz in the domain wall coordinates.  Then  we present the equations  of motion in different configuration  spaces.
In Section \ref{app:AppBM} we construct a near-horizon asymptotically AdS$_{3}$ black hole from the dynamical system and discuss its thermodynamical properties. In Section \ref{sec:solPhiImp} we find of a special class of RG flows with the condition that the scalar field slowly changes. Such solution  can be represented analytically from  the horizon to the boundary. We also explore the behaviour of the solution for the  scalar field of the boundary and show that it is in  agreement with known results. In Section \ref{sec:discussion} we give a discussion of the results and future directions.

Some of the computations in this article have been performed by means of the computer algebra system SageMath \cite{sagemath}; the
relevant notebook is posted at\\
\url{https://nbviewer.org/url/relativite.obspm.fr/notebooks/BH_3D_supergravity.ipynb}

\setcounter{equation}{0}

\section{Setup}\label{sec:setup}
\subsection{The holographic model}
 The  action of the $3D$ truncated supergravity reads
\be\label{act}
S = \frac{1}{16\pi G_{3}} \int d^{3}x\sqrt{|g|}\left(R -\frac{1}{a^2}(\partial\phi)^2 -V(\phi)\right) + \frac{1}{8\pi G_{3}}\int_{\partial M} d^2 x \sqrt{|\gamma|}K,
\ee
with the potential $V(\phi)$  is given by:
\be\label{pot}
V(\phi)=2\Lambda_{\rm uv} \cosh^2\phi\left[(1-2a^2)\cosh^2\phi+2a^2\right],
\ee
where the parameter $a$ is non-zero and  is related to the curvature of the target manifold. In what follows we focus on the cases: {\bf a)} $a^2\leq\frac{1}{2}$ and {\bf b)}  $\frac{1}{2}\leq a^2\leq 1$.

 The Gibbons-Hawking-York (GHY) boundary term  in (\ref{act}) includes the determinant of the  induced metric  $\gamma$ on the boundary $\partial M$ and the extrinsic curvature $K$
 \be
 K =\gamma^{\mu\nu}K_{\mu\nu}, \quad K_{\mu\nu} =-\nabla_{\mu}n_{\nu}=\frac{1}{2}n^{\rho}\partial_{\rho}\gamma_{\mu\nu}, \quad n^{\mu}=\frac{\delta^{\mu}_{w}}{\sqrt{g_{ww}}}.
 \ee

\begin{figure}[h!]
    \centering
    \includegraphics[width= 7cm]{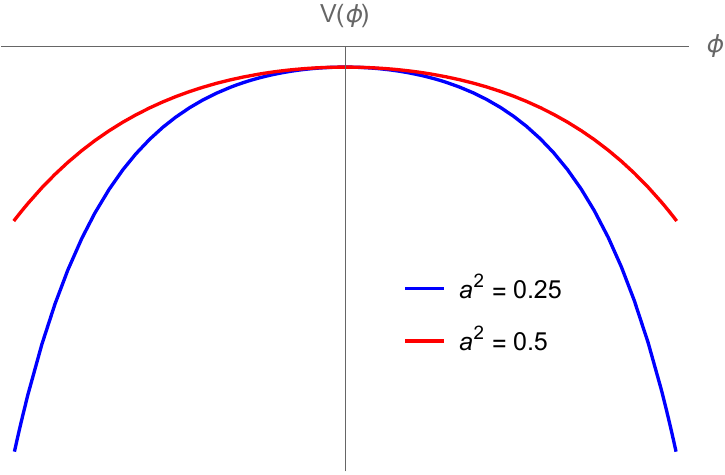} \quad \includegraphics[width= 7cm]{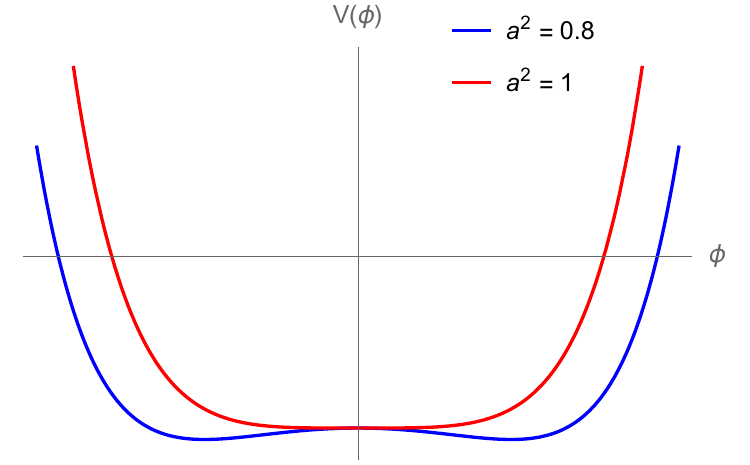}
    \caption{\small The potential (\ref{pot}) against the scalar field $\phi$ for different values of $a$. For all we set $\Lambda_{\rm uv}=-1$.  }
    \label{fig:scpotential}
\end{figure}

  We show the behaviour of the potential on the scalar field in Fig.~\ref{fig:scpotential}.  Note that for $a^2\leq \frac{1}{2}$ and for $a^2\geq 1$ the potential has one extremum,  namely,
  \be\label{phi10}
  \phi_{1} =0,
  \ee while for $\frac{1}{2}<a^2<1$ the  potential has two more extremal points, i.e.
 \be \label{crP-v}
 \phi_{2,3} = \frac{1}{2}\ln\left(\frac{1\pm 2|a|\sqrt{1-a^2}}{2a^2 -1}\right).
 \ee
Moreover, if we consider the case $a^2>\frac{1}{2}$ the scalar potential has also zeroes:
\be\label{scpV0}
\phi_{\pm} = \pm \cosh ^{-1}\left(\frac{a}{\sqrt{a^2-\frac{1}{2}}}\right).
\ee
The behavior of the potential $V$ on $\phi$ for various $a$ is presented in Figs.~\ref{fig:scpotential}. We marked by $\phi_{3}$  and $\phi_2$ (\ref{crP-v}) the left and right extrema of the potential, correspondingly.

%The generic form  of the stress-energy tensor for  the action (\ref{act}) can be represented as
%\be\label{STE}
%T_{\mu\nu}=\frac{1}{a^2}\left(\partial_{\mu}\phi\partial_{\nu}\phi-\frac{1}{2}g_{\mu\nu}\partial_{\sigma}\phi\partial^{\sigma}\phi\right)-\frac{1}{2}g_{\mu\nu}V.
%\ee

We focus on holographic RG flows at finite temperature for the model (\ref{act}) for these reasons  we consider the following ansatz of the  metric in the domain wall coordinates
\be\label{metricmain}
ds^2 = e^{2A(w)}\left(-f(w)dt^2+dx^{2}\right) +\frac{dw^2}{f(w)},
\ee
with the radial coordinate running the region $w\in (w_{h},+ \infty)$, such that $w_{h}$ is a position  of the horizon  and the conformal boundary is located at $w\to +\infty$. In (\ref{metricmain})  the scale factor  $A=A(w)$, the function and $f(w)$ defines  the position of the horizon, such  that it has the following structure
\be\label{horizonfunc}
 \begin{cases}
     f<0, & \text{for $w< w_h$},\\
     f=0,&\text{for $w=w_h$},\\
    f>0, & \text{for $w> w_{h}$},
  \end{cases}
\ee
and $f=1$ on the boundary of the spacetime.
The temperature of the solution (\ref{metricmain}) can be found as  \cite{Gursoy:2018umf}

\be\label{HTemp}
T_{H} = \left. \frac{e^{A(w_{h})}}{4\pi}\frac{df}{dw} \right| _{w= w_{h}}.
\ee

The Hawking temperature (\ref{HTemp}) of the black hole  (\ref{metricmain}) corresponds to the temperature of the dual field theory.

The entropy density of the solution is given by
\be\label{endDens}
s = 4\pi M_{p}e^{A(w_{h})},
\ee
with
\be
M_{p} =\frac{1}{16\pi G_{3}}.
\ee

 The scalar field profile has   the dependence only on the radial coordinate $w$
\be
\phi = \phi(w).
\ee

We are looking for  black hole solutions with the metric  (\ref{metricmain}) such that the scalar field has some finite value at the horizon $w_{h}$
\be\label{phiconst}
\phi(w_{h}) = \phi_{h}.
\ee
%Holographic RG flows and its  fixed points for the gravity model (\ref{act})-(\ref{pot}) with  an ansatz of the metric without a black hole, i.e. for (\ref{metricmain}) with $f=1$, was studied in \cite{Golubtsova:2022hfk,NE-RG}.

In what follows it is convenient to introduce the function
\be\label{funcgmain}
g = \ln f.
\ee

Then the Einstein equations to (\ref{act}) can be represented in the following form
\bea
\ddot{A}+\frac{\dot{\phi}^2}{a^2}&=&0,\label{eom1}\\
\ddot{g}+\dot{g}^2+2\dot{g}\dot{A}&=&0,\label{eom2}\\
\dot{A}\dot{g}+2\dot{A}^2-\frac{\dot{\phi}^2}{a^2}+e^{-g}V&=&0,\label{eom3}
\eea
where the overdot indicates the derivative with respect the radial coordinate $w$.
 The  scalar field equation is
\bea\label{eqd}
\ddot{\phi}+\dot{g}\dot{\phi}+2\dot{A}\dot{\phi}- \frac{a^2}{2}e^{-g}V_{\phi}=0,
\eea
where $V_{\phi}$ is a derivative of the potential with respect to $\phi$.

Note, that for $f=1$ the model has an exact half-supersymmetric solution, which interpolates between AdS and scaling geometries \cite{Deger:2002hv}.

\subsection{The dynamical system in $\mathbf{R}^3$}

 We are looking black hole solutions (\ref{metricmain}) with (\ref{phiconst}) to (\ref{act}), which can be interpreted as holographic RG flows at finite temperature. For this, we represent eqs.(\ref{eom1})-(\ref{eqd}) as an autonomous dynamical system
 introducing new variables \cite{Gursoy:2008za,Arefeva:2018jyu}
\be\label{XvarDef}
X=\frac{d\phi}{d A} = \frac{\dot{\phi}}{\dot{A}},\qquad Y=\frac{d g}{d A}=\frac{\dot{g}}{\dot{A}},
\ee
where $g=\ln f$,  $\phi$ is the dilaton, $A$ is the scale factor, and the derivatives are taken with respect to the radial coordinate $w$. So both variables $X$ and $Y$ run from $-\infty$ to $\infty$. In \cite{Gursoy:2018umf} it was shown that the first order formalisms for the holographic RG flows in terms of the  the  variables $X,Y$ and the superpotential are equivalent.

As for the scalar field it is convenient to compactify it using the following redefinition:
\be\label{Zvar}
z = \frac{1}{1+e^{\phi}},
\ee
so that $z\in[0,1]$ for $\phi \in(-\infty;\infty)$.
Below we will use (\ref{crP-v}) and (\ref{scpV0}) as  $z_{i}=z(\phi_{i})$ with $i=1,2,3$ and $z_{\pm}=z(\phi_{\pm})$.

The ratio of the potential and its derivative  in terms of the variable $z$ (\ref{Zvar})  is represented as follows
\be\label{VphV}
\frac{V_{\phi}}{V}= -\frac{4(4a^2z^4-8a^2z^3+12a^2z^2-4z^4-8a^2z+8z^3+2a^2-8z^2+4z-1)(2z-1)}{(8a^2z^2-4z^4-8a^2z+8z^3+2a^2-8z^2+4z-1)(2z^2-2z+1)}.
\ee

Taking into account (\ref{XvarDef})-(\ref{Zvar}) the equations of motion (\ref{eom1})-(\ref{eqd})  are brought to
\bea\label{dz/da}
&&\frac{dz}{dA}=z(z-1)X,\\ \label{dx/da}
&&\frac{dX}{dA}=\left(\frac{X^2}{a^2}-Y-2\right)\left(X+\frac{a^2}{2}\frac{V_{\phi}}{V} \right),\\ \label{dy/da}
&&\frac{dY}{dA}=Y\left(\frac{X^2}{a^2}-Y-2\right),
\eea
where in  (\ref{dx/da}) we also suppose (\ref{VphV}).

 The equilibrium points  of the dynamical system (\ref{dz/da})-(\ref{dy/da})  can be related with the  fixed points of holographic RG flows. It worth to be noted that the critical points with $Y=0$ correspond to zero-temperature regime.
 Thus, the conformal boundary of the black hole spacetime corresponds to $Y\to 0$,
while at the horizon $Y$ tends to infinity $Y\rightarrow \infty$.

%The corresponding geometries include both AdS and scaling spacetimes. However, only for $a^2\leq \frac{1}{2}$ the solutions with scaling obey  Gubser's bound (i.e. the curvature singularity is good according to the Gubser classification).

%We are interested in  fixed points of the dynamical system, which correspond to  near-horizon black hole solutions, i.e. $Y\to \infty$.

% Below we apply Poincar\'e transformations to project
 % the 3d dynamical system (\ref{dz/da})-(\ref{dy/da}) defined on $\mathbf{R}^3$ into the unit cylinder.
 % This helps us to explore infinite points corresponding to near-horizon black hole solutions and explore the global phase portrait of (\ref{dz/da})-(\ref{dy/da}).

\subsection{The dynamical system in the  cylinder}\label{sec:cylmap}

To understand the behaviour of flows at infinity we apply the so-called Poincar\'e projection \cite{10.5555/102732}, which maps the system into the unit cylinder.
 We use Poincar\'e projection only over the variables $X$ and $Y$ due to  $z$ is compact.

To extract the information about the finite temperature flows we map  the system (\ref{dx/da})-(\ref{dy/da}) defined on the plane $\mathbf{R}^2$  into the disc $\mathbf{D}^2$.
This allows to consider infinite points of eqs.(\ref{dx/da})-(\ref{dy/da}), which are  difficult to be caught considering the system on $\mathbf{R}^2$.

The coordinates of the system on  $\mathbf{R}^2$ are related  with  the coordinates on $\mathbf{D}^2$ by
\be \label{ch-cor}
 X=\frac{x}{\sqrt{1-x^2-y^2}} \,, \quad Y=\frac{y}{\sqrt{1-x^2-y^2}}.
\ee
For the new coordinates $x, y$ on  $\mathbf{D}^2$ we also have the constraint
\be \label{cyl-1}
x^2+y^2 \leq 1.
\ee

Applying the coordinate transformations \eqref{ch-cor} for $X$ and $Y$ to  eqs. (\ref{dz/da})-(\ref{dy/da}), we come to the dynamical system in the $3d$ unit cylinder
\bea \label{EqsonCyl001}
z' &= &z (z-1) x \, ,\\
 \label{EqsonCyl0012}
       x' &= &\mathrm{p}(x,y,z),  \\
\label{EqsonCyl0013}
           y'&= &\mathrm{q}(x,y,z) ,
\eea
where $\mathrm{p}$, $\mathrm{q}$ are represented by

\bea \nonumber
\mathrm{p} &=&  \left( \sqrt{1-x^2-y^2} \left(2 \sqrt{1-x^2-y^2}+y\right)-\frac{x^2}{a^2} \right) \left(\frac{a^2}{2}\frac{V_{\phi}}{V}(x^2-1) -x \sqrt{1-x^2-y^2}\right) \,, \\
\label{q-disk2}
\mathrm{q} &=&  \left( \sqrt{1-x^2-y^2} \left(2 \sqrt{1-x^2-y^2}+y\right)-\frac{x^2}{a^2} \right)  \left(\frac{a^2}{2}\frac{V_{\phi}}{V} x -\sqrt{1-x^2-y^2}\right) y
\eea
and the derivatives are redefined as
\be \label{derred}
\chi' = \sqrt{1-x^2-y^2}\, d\chi/dA.
\ee

This  is admissible, since, such redefinition does preserves the behavior of  flows, and at the same time allows us to avoid divergences, which arise for $x^2+y^2=1$.

The dynamical system (\ref{EqsonCyl001})-(\ref{EqsonCyl0013}) with (\ref{q-disk2}) has a rich structure of equilibrium sets, that was investigated in \cite{Golubtsova:2024dad}.  We omit the discussion of this issue  here, keeping in mind only, that  the  equilibrium points $p_{i}$  have the coordinates $(0,0,z_{i})$, where  $z_{i}$ takes the value  at the $i$-extremum, and correspond to  the AdS spacetimes, while the points $\bar{p}_{i}$ are related with the near-horizon regions and have the coordinates  $(0,1,z_{i})$.
%\textcolor{red}{It is also worth to be noted that  for $\frac{1}{2}<a^2<1$ we have  the special lines $Q_{+}Q'_{+}$, $Q_{-}Q'_{-}$, which are related with the points, where the potential  vanishes.}

\begin{figure}[t]
    \centering
  \includegraphics[width=10cm]{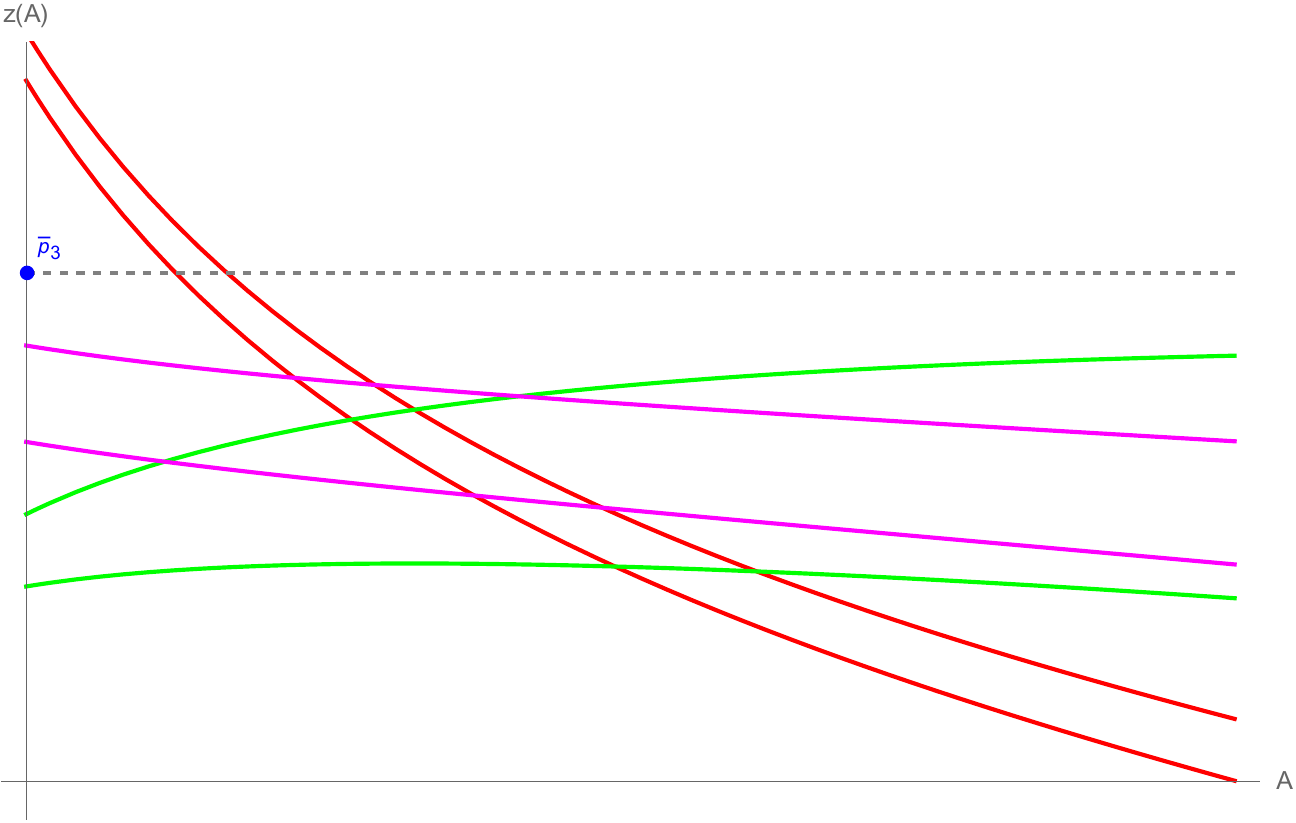}
    \caption{Three different types of numerical trajectories  near the  horizon  for the component $z(A)$ of the dynamical system (\ref{dz/da})-(\ref{dy/da}) with initial conditions (\ref{icC4}), $a^2=0.8$. All trajectories  correspond to the asymptotically AdS black holes in the near-horizon region. For red: $\upsilon > 0, \, \delta>0$. For magenta: $\upsilon<0,\, \delta<0$. For green (non-monotonic): $\upsilon>0, \, \delta<0$. The dark grey dashed line is the exact AdS  black hole solution \eqref{BTZbh}-\eqref{BTZbh2} with $\phi_h = \phi_3$.}
    \label{Fig:cylz23zpm}
\end{figure}

The asymptotically AdS black hole solutions  to eqs.(\ref{EqsonCyl001})-(\ref{EqsonCyl0013}) were found numerically in \cite{Golubtsova:2024dad}, such that for $a^2\leq\frac{1}{2}$ we used the following initial conditions
\be\label{icC2}
x=0,\;\;\; y=1-\varepsilon,\;\;\; z\in(0,1),
\ee
i.e. the value of the scalar field on the horizon $\phi_h$ can take any value of the range of the potential, the initial value of $y$ should be close to the near horizon region and, finally, $x$ and $y$ should respect (\ref{cyl-1}).  For the case  $\frac{1}{2}<a^2<1$ the initial  conditions are slightly modified such that $\phi_h$ runs either in the  interval $(\phi_{2},\phi_{1})$ or  $(\phi_{1},\phi_{3})$, i.e.
\be \label{icC3}
x=0,\;\;\; y=1-\varepsilon,\;\;\; z\in[z_2+\delta,z_3-\delta] \,,\ee
where $\delta > 0$.
For $\frac{1}{2}<a^2<1$  there is a broader class of asymptotically AdS black hole solutions with the initial conditions given by
\be\label{icC4}
x=\upsilon,\;\;\; y=1-\varepsilon,\;\;\;  z\in[z_2-\delta,z_3+\delta],
\ee
with $\upsilon\neq0$, such that for $\phi_h\in(\phi_{3},\phi_{3} + \delta)$ the parameter $\upsilon$ is positive and for $\phi_h\in (\phi_{2},\phi_{2}-\delta)$ the $\upsilon$ is negative. The dynamics of the  solutions with (\ref{icC2}),(\ref{icC3}) and  (\ref{icC4}) was analyzed in \cite{Golubtsova:2024dad}.
Note, that the some solutions with the initial conditions (\ref{icC4}) have  a non-monotonic behaviour of $\phi$ comparing to those with \eqref{icC2} or \eqref{icC3}. In  Fig.~\ref{Fig:cylz23zpm}  we plot   $z$ of  the numerical black hole solutions as  a function  of $A$  with the initial conditions \eqref{icC4} for the case of three extrema of the potential with  $a^2=0.8$. We show the behaviour of the solutions in the near-horizon region. The trajectories start near the  point $\bar{p}_{3}$, which is related with $\phi_{3}$,  and end in $p_{1}$ corresponding to $\phi_{1}$.

Below we will present analytic description of black hole solutions, which are related with (\ref{icC2}) and (\ref{icC3}), and discuss its thermodynamical properties.

\setcounter{equation}{0}

\section{Asymptotically AdS (BTZ) black hole solutions}\label{app:AppBM}

\subsection{Exact BTZ black holes } \label{app:AppB}

AdS black holes (non-rotating BTZ solution)  can be obtained as a solution of the autonomous system. It corresponds to $X=0$ and $Z=const$ in
(\ref{XvarDef})-(\ref{Zvar}) and $\phi_h=\phi_{i}$, $i=1,2,3$. The metric of
the 3d AdS black hole in the domain wall coordinates reads
\be\label{BTZbh}
ds^2=e^{2A}\left(-f(w)dt^2 +dx^2\right)+\frac{dw^2}{f(w)},
\ee
where
\be\label{BTZbh2}
A = \sqrt{-\frac{V(\phi_{h})}{2}} w, \quad f = 1 -e^{-\sqrt{-2V(\phi_h)}(w-w_h)}.
\ee
The boundary of the spacetime is reached  as $w\to\infty$  and the horizon  $w\to w_h$.

Setting $x=\varphi$ with  $\varphi\in[0,2\pi)$, the metric (\ref{BTZbh})-(\ref{BTZbh2}) takes a standard form of the non-rotating BTZ black hole \cite{Kraus:2006wn}.

The temperature and the entropy of the BTZ black hole solution are
\be\label{conftherm}
T_H  =\frac{1}{2\pi\ell}e^{w_h/\ell},\quad s=4\pi M_{p}e^{w_h/\ell}.
\ee

In the near-horizon region the metric functions takes the form (\ref{BTZbh})
\be\label{nhBTZ}
A|_{w\to w_h}\simeq \sqrt{-\frac{V(\phi_h)}{2}}w_h, \quad f|_{w\to w_h} =\sqrt{-2V(\phi_h)}(w-w_h)+\ldots\,.
\ee

 \subsection{Near-horizon asymptotically AdS black holes}\label{sec:nhsol}
In \cite{Golubtsova:2024dad}  it was shown that  the equation for the variable $Y$ of the system (\ref{dz/da})-(\ref{dy/da}) can be split off for the near-horizon region with  $Y\to\infty$ and easily integrated
\be
Y = \frac{1}{A-A_{h}},
\ee
where  $A_{h}=A(w_h)$. Now it is convenient to find a solution for $\phi$ in terms of  $A$ and then find  $A$.
The other two equations can be represented as a single second order equation for  $\phi$
\be \label{one-eq-XF-1}
(A-A_{h})\frac{d^2 \phi}{dA^2}+\frac{d\phi}{dA}+\frac{a^2}{2} \frac{V_{\phi}(\phi)}{V(\phi)}=0.
\ee
To proceed we expand the term $\frac{V_{\phi}(\phi)}{V(\phi)}$ in Taylor series near $\phi_h$ % up to the  second order
\bea
\label{ppp-ex}
\frac{a^2}{2}\frac{V_{\phi}(\phi)}{V(\phi)}\Big|_{\phi_{h}}
%&\cong & \frac{a^2}{2}\left(\frac{V_{\phi}(\phi_{h})}{V(\phi_{h})} + \left(\frac{V_{\phi\phi}(\phi_{h})}{V(\phi_{h})}-\frac{V_{\phi}(\phi_{h})^2}{V(\phi_{h})^2}\right) (\phi -\phi_{h})\right)\\
&=&\mathsf{\Lambda}^{(h)}+\mathsf{K}^{(h)}(\phi-\phi_{h}),
\eea
where we denote
\be \label{SlDef}
\mathsf{\Lambda}^{(h)}=\displaystyle{\frac{a^2}{2}\frac{V_{\phi}(\phi_h)}{V(\phi_h)}},\quad  \Delta^{(h)} =\frac{a^2}{2}\frac{V_{\phi\phi}(\phi_{h})}{V(\phi_{h})},\quad\mathsf{K}^{(h)}=\left(\Delta^{(h)}-\frac{2}{a^2}\left(\mathsf{\Lambda}^{(h)}\right)^2\right).
\ee
%\bea\label{solDilStar}
%\phi(A) =
%\begin{cases}
%\mathrm{c_1} J_0\left(2 \sqrt{\mathsf{K}^{(h)}(A-A_{h}) }\right)+ \mathrm{c_2} Y_0\left(2 \sqrt{\mathsf{K}^{(h)}(A-A_{h})}\right)+\phi_{h} -\frac{\mathsf{\Lambda}^{(h)} }{\mathsf{K}^{(h)}},&\text{for}\quad \mathsf{K}^{(h)}>0, \\
%\,\\
%\mathrm{c_1} I_0\left(2 \sqrt{|\mathsf{K}^{(h)}(A-A_{h})|}\right)+ \mathrm{c_2} K_0\left(2 \sqrt{|\mathsf{K}^{(h)}(A-A_{h})| }\right)+\phi_{h} -\frac{\mathsf{\Lambda}^{(h)} }{\mathsf{K}^{(h)}},&\text{for}\quad \mathsf{K}^{(h)}<0\,,
%\end{cases}
%\eea
%where $I_{0}$ and $K_{0}$ are modified Bessel functions of the first and second kinds, correspondingly.

Note, that for $a^2\leq \frac{1}{2}$ the quantity $\mathsf{K}^{(h)}$ is  always non-negative, while for $\frac{1}{2}<a^2<1$ we have to trace the sign of  $\mathsf{K}^{(h)}$ with respect to the position of $\phi_{h}$ and also take into account zeros of the scalar potential.

The solution to eq. \eqref{one-eq-XF-1} with (\ref{ppp-ex}) near $A_{h}$  can be  represented as follows
\bea \label{gen-phi}
\phi(A) = \phi_{h}-\mathsf{\Lambda}^{(h)}(A-A_{h}),
\eea
with $\mathsf{\Lambda}^{(h)}$ given by (\ref{SlDef}). Note that the solution \eqref{gen-phi} corresponds to the following initial conditions
\be \label{mi-1}
\phi(A_h) = \phi_h \,, \quad \phi'(A_h) = - \mathsf{\Lambda}^{(h)} \,.
\ee
%The expansion in series of the scalar field (\ref{solDilStar})   up to the first order
Subtracting (\ref{eom1}) from (\ref{eom3}) we come the equation
\be\label{mastEqA-2}
\ddot{A} + \dot{A}^2\left(\frac{d\phi(A)}{dA}\right)^2 = 0.
\ee

Taking into account (\ref{gen-phi}), we find  a solution for the scale factor
\be \label{fsolA}
A(w) = \frac{a^2}{(\mathsf{\Lambda}^{(h)})^2} \ln \left(\frac{\frac{(\mathsf{\Lambda}^{(h)})^2}{a^2} w+\mathrm{c_{A}}}{\mathrm{c_{A}}}\right) \,,
\ee
with
\be\label{cAm}
\mathrm{c_{A}} = \sqrt{-\frac{2}{V(\phi_{h})}},
\ee
where the constants of integration are chosen such that (\ref{fsolA}) matches with the near-horizon AdS black hole solution if $\phi_{h}$ is the critical point of the potential.

The blackening function $f$ near the horizon us
\be \label{forGA}
f = e^{\mathrm{c_g}}(A-A_h),
\ee
with the integration constant $\mathrm{c_g}$, which reads from the which is found from  the Einstein equations as follows
\be \label{CGE}
\mathrm{c_g} = \ln \left (|V(\phi_h)|\left(\mathrm{c_A}+\frac{(\mathsf{\Lambda}^{(h)})^2}{a^2} w_h\right)^2 \right ).
\ee

%The near horizon geometry of the black hole is given by
%\be\label{solmetrF}
%ds^2 \simeq  \left(\frac{\frac{(\mathsf{\Lambda}^{(h)})^2}{a^2} w+\mathrm{c_{A}}}{\mathrm{c_{A}}}\right)^{\frac{2a^2}{(\mathsf{\Lambda}^{(h)})^2}}\left(-f dt^2 +dx^2\right)+\frac{dw^2}{f},
%\ee
%with $f$ given by
%\be\label{fsolmain}
%f =  \frac{a^2 e^{\mathrm{c_{g}}}}{(\mathsf{\Lambda}^{(h)})^2} \ln \left(\frac{\frac{(\mathsf{\Lambda}^{(h)})^2}{a^2} w+\mathrm{c_{A}}}{\frac{(\mathsf{\Lambda}^{(h)})^2}{a^2} w_{h}+\mathrm{c_{A}}}\right) \,, \ee
%with $\mathrm{c_{g}}$  defined by (\ref{CGE}). The scalar field of the solution is defined by (\ref{gen-phi}) with (\ref{genconst}).

 Introducing the quantity
\be\label{kappapar}
\kappa = \frac{(\mathsf{\Lambda}^{(h)})^2}{a^2},%\quad  \mathsf{\Lambda}^{(h)}=\displaystyle{\frac{a^2}{2}\frac{V_{\phi}(\phi_h)}{V(\phi_h)}}.
\ee
where the parameter  $\mathsf{\Lambda}^{(h)}$ is given by (\ref{SlDef}),
  the black hole metric  can be represented as follows
\be\label{simpBHm}
ds^2 \simeq \left(1+\frac{\kappa}{\mathrm{c}_{A}}w\right)^{2/\kappa}\left(- fdt^2 +dx^2\right) +\frac{dw^2}{f},
\ee
with
the blackening function (\ref{forGA}) reads as
\bea\label{blackfm}
f &=& \frac{e^{\mathrm{c}_{g}}}{\kappa}\ln\left(\frac{1+\frac{\kappa}{\mathrm{c}_{A}}w}{1+\frac{\kappa}{\mathrm{c}_{A}}w_h}\right)
%&=&e^{\mathrm{c}_{g}}\left(A-A_{h}\right)\simeq-\frac{2}{\mathsf{\Lambda}^{(h)}}\left(\phi -\phi_{h}\right)=-\frac{4}{a^2}\frac{V(\phi_{h})}{V'(\phi_h)}(\phi-\phi_h),
\eea
with $\mathrm{c}_{A}$ defined by (\ref{cAm}) and $\mathrm{c}_{g}$ (\ref{CGE}) represented as
\be
\mathrm{c}_{g} =\ln\left(2\left(1+\frac{\kappa}{\mathrm{c}_{A}}w_h\right)^2\right).
\ee
From the latter one can see that for  $\phi_{h}=\phi_{1,2,3}$,  $\mathrm{c_g}$  turns to be $\mathrm{c_{g}}=\ln2$ as for the near-horizon metric of the AdS black hole.

The scalar field of the solution (\ref{gen-phi}) is  non-zero and given by
\be\label{scfsol}
\phi(A) = \phi_{h}-\frac{\mathsf{\Lambda}^{(h)}}{\kappa}
\ln \left(\frac{1+\frac{\kappa}{\mathrm{c_{A}}}w}{1+\frac{\kappa}{\mathrm{c_{A}}}w_h}\right).
\ee
For vanishing   $\mathsf{\Lambda}^{(h)}$, corresponding to extrema of the potential $V_{\phi}=0$, the near-horizon  solution  turns to be the BTZ black hole with a constant scalar field  and near-horizon metric functions \eqref{nhBTZ}.

\subsection{Thermodynamics of the near-horizon  solutions}

Here we discuss the thermodynamics of the black hole solutions, which we constructed in the previous section. We note, that in the case of one extremum of the potential, $a^2\leq \frac{1}{2}$, the  value of the scalar field at the horizon $\phi_h$ can take any value on the  range of $V$.  For the case of the  potential with  three  extrema, $\frac{1}{2}<a^2<1$ we are restricted as $\phi_{h}\in (\phi_{2},\phi_{3})$.

\begin{figure}[t]
    \centering
     \includegraphics[width=7cm]{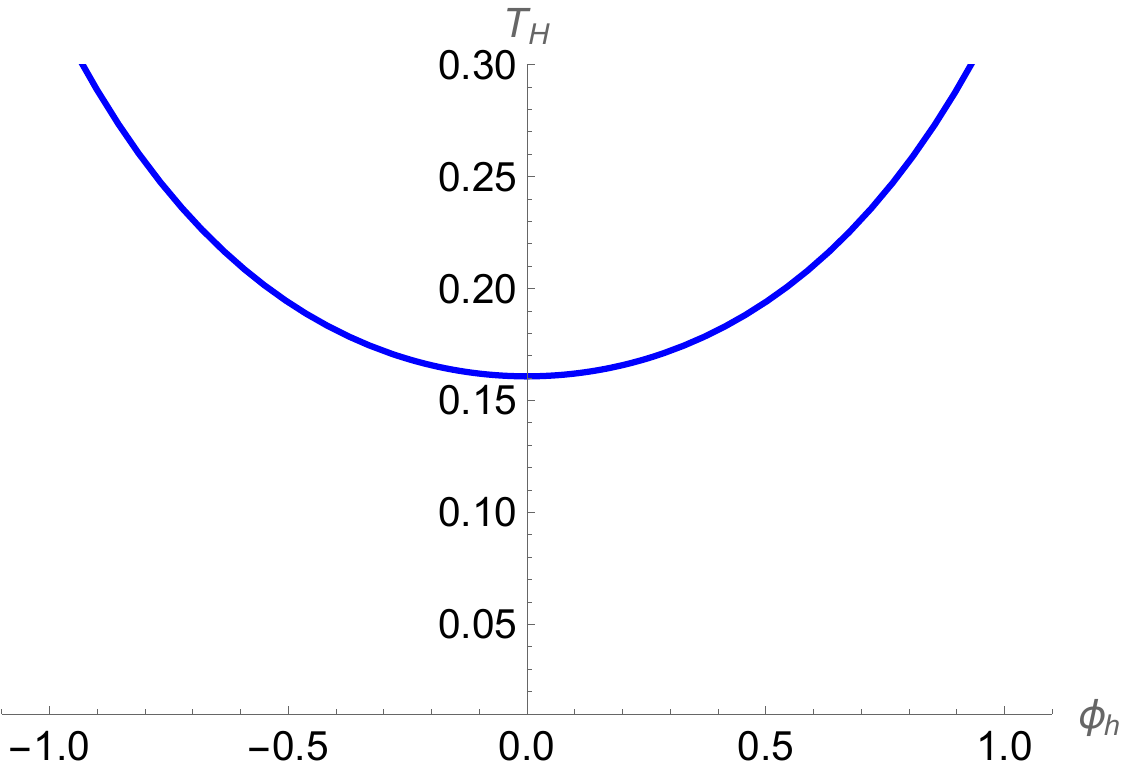}{\bf a)}
     \includegraphics[width=7cm]{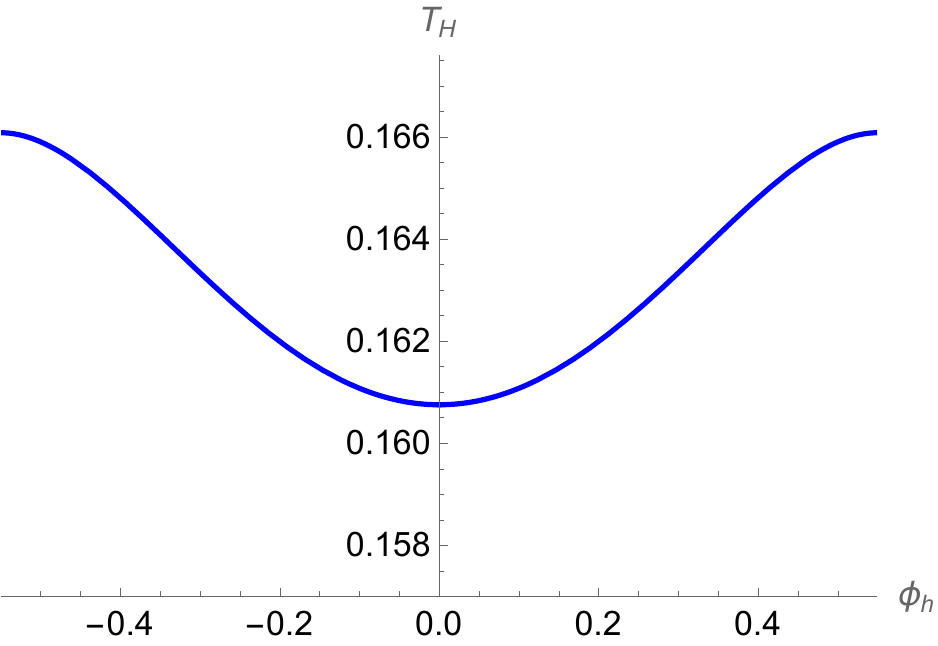}{\bf b)}
     \caption{\small Hawking temperature (\ref{bhtemp}) as a function of  $\phi_h$: a) $a^2 =0.25$, b) $a^2=0.8$; for all we set $w_{h}=0.01$. }
    \label{Fig: HawkingT02508}
\end{figure}

The Hawking temperature of the black hole solution (\ref{simpBHm}) with (\ref{blackfm}) and (\ref{scfsol}) can be calculated using (\ref{HTemp}).
Then the temperature reads
\be\label{bhtemp}
T_{H} =
%\frac{1}{2\pi\mathrm{c}_{A}}\left(1+\frac{\kappa}{\mathrm{c}_{A}}w_h\right)^{\frac{\kappa+1}{\kappa}}=
\frac{1}{2\pi \mathrm{c}_{A}}\mathsf{B}^{\frac{\kappa+1}{\kappa}},
\ee
where we introduced the quantity
\be
\mathsf{B}= 1+\frac{\kappa}{\mathrm{c}_{A}}w_h,
\ee
where $\mathrm{c}_{A}$ is given by  (\ref{cAm}) and $\kappa$ by (\ref{kappapar}).
We present the behaviour of the temperature as a function of the value of the scalar field on the horizon $\phi_h$ in Fig.~\ref{Fig: HawkingT02508} for $a^2=0.25$ and $a^2=0.8$ with fixed $w_h$. The temperature symmetrically increases  for opposite values of $\phi_h$, for $a^2=0.8$ the temperature is bounded by a maximal value due to the restriction for the value of $\phi_h$ .
At the extrema of the potentials , that corresponds to $\kappa\to 0$ \eqref{kappapar}, the  thermodynamics becomes conformal and the temperature is given by \eqref{conftherm} for both cases of $a^2$. This can be  checked expanding in series  (\ref{bhtemp}) by small $\mathsf{\Lambda}^{(h)}$.
Thus, near the extremum of the potential $V$ with $\mathsf{\Lambda}^{(h)}\simeq 0$ the temperature reads
\be\label{tempexp}
T_{H,\mathsf{\Lambda}^{(h)}\to 0}=\frac{e^{w_h/\mathrm{c}_A}}{2\pi \mathrm{c}_A}+\frac{e^{w_h/\mathrm{c}_A}(2\mathrm{c}_A-w_h)w_h(\mathsf{\Lambda}^{(h)})^2}{4\pi a^2\mathrm{c}^{3}_A}.
\ee

\begin{figure}[h!]
    \centering
     \includegraphics[width=7cm]{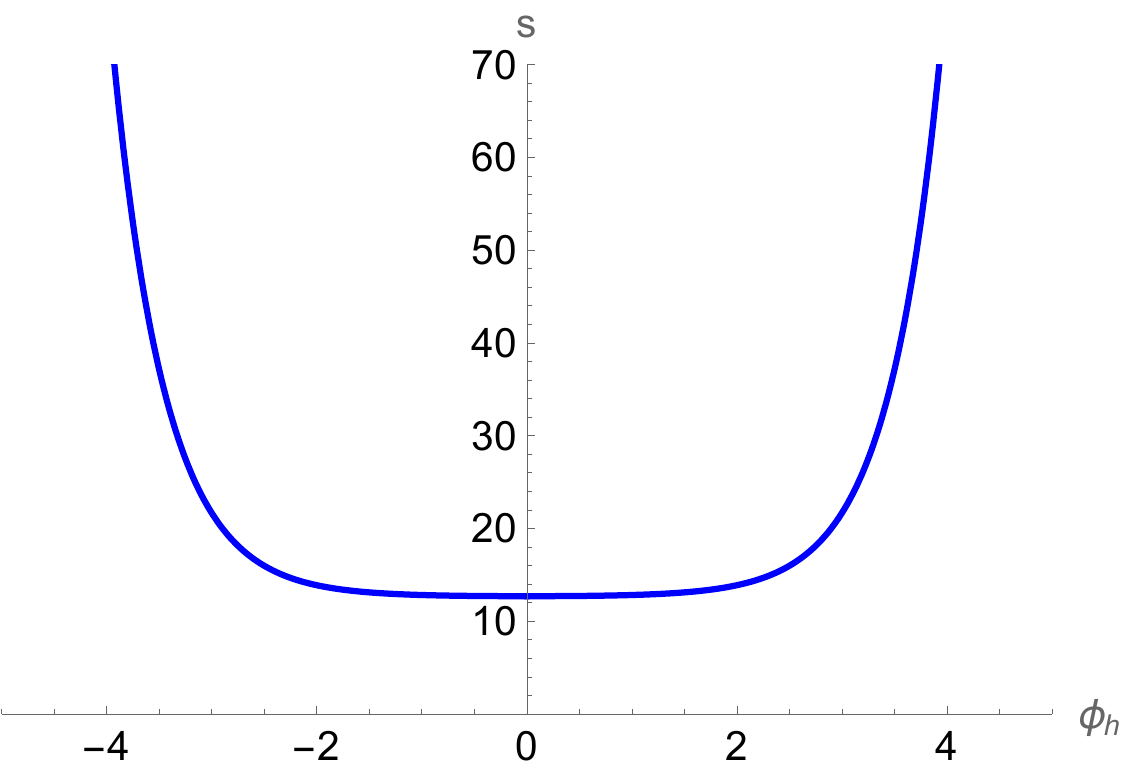}{\bf a)}\,\,\,\,
     \includegraphics[width=7cm]{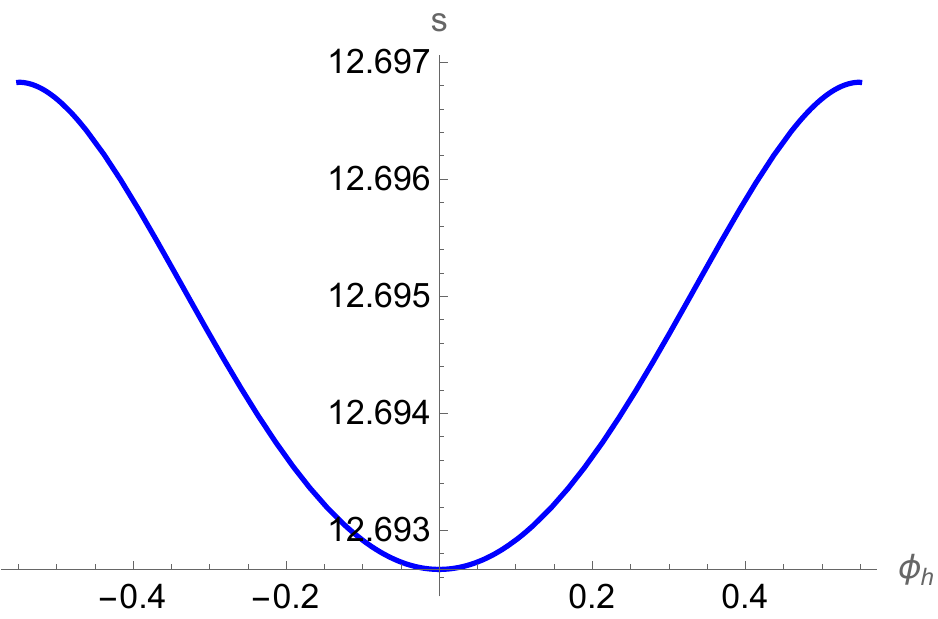}{\bf b)}
     \caption{\small Entropy density (\ref{entbh}) as a function of  $\phi_h$: a) $a^2 =0.25$, b) $a^2=0.8$; for all we set $w_{h}=0.01$. }
    \label{Fig: Entropy02508}
\end{figure}

Using (\ref{endDens}) one finds that the entropy of the black hole solution (\ref{simpBHm}) with (\ref{blackfm})  is given by
\be\label{entbh}
s= %4\pi M_{p}\left(1+\frac{\kappa}{\mathrm{c_{A}}}w_h\right)^{\frac{1}{\kappa}}=
4\pi M_{p}\mathsf{B}^{\frac{1}{\kappa}}\,.
\ee
We show the behavior of the entropy (\ref{entbh}) as a function of $\phi_h$ in Fig.~\ref{Fig: Entropy02508}. One can see that for $a^2=0.8$ the entropy has its  maximal value  comparing to the case $a^2=0.25$, for which the entropy is unbounded from above.
Again, if $\phi_h$ matches with the value at extemum the entropy is given by  (\ref{conftherm}).

\begin{figure}[t]
    \centering
     \includegraphics[width=6cm]{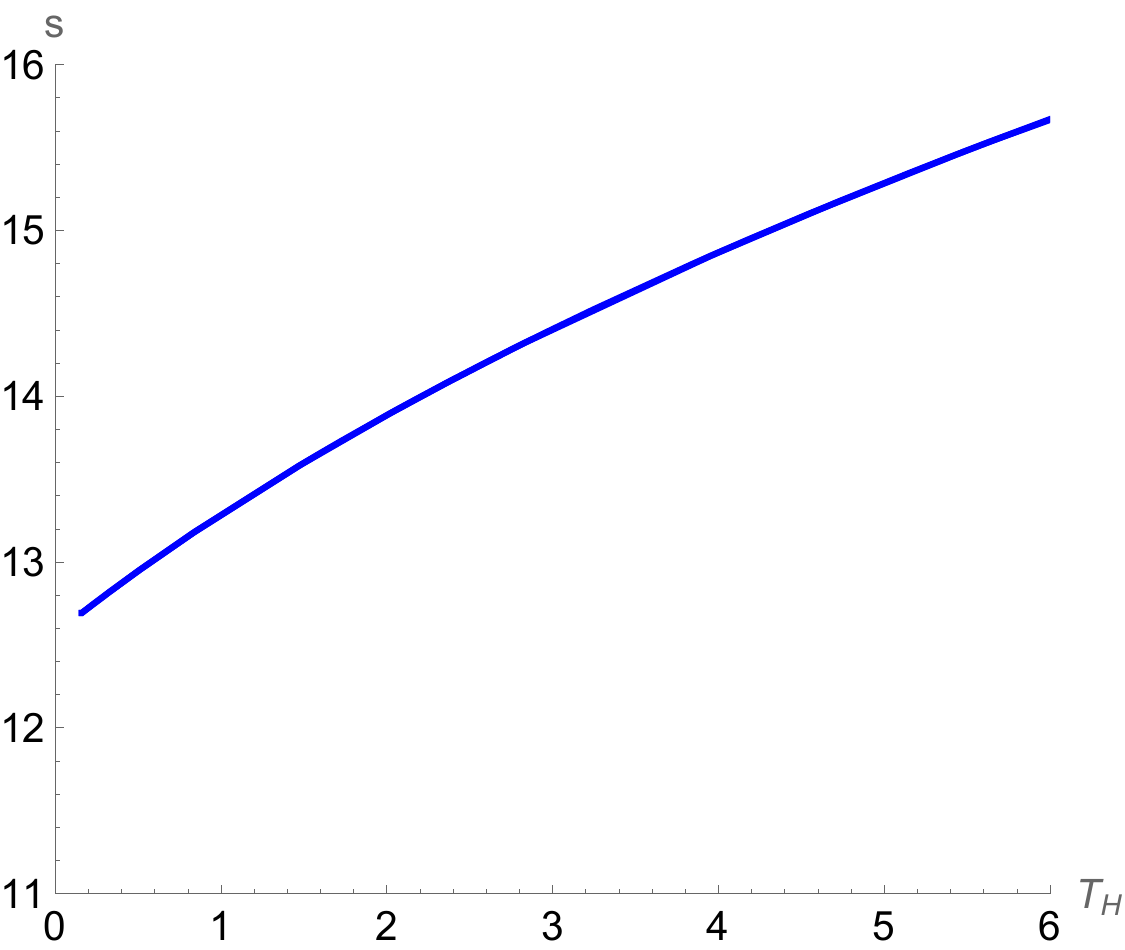}{\bf a)}\,\,\,\,
     \includegraphics[width=6.5cm]{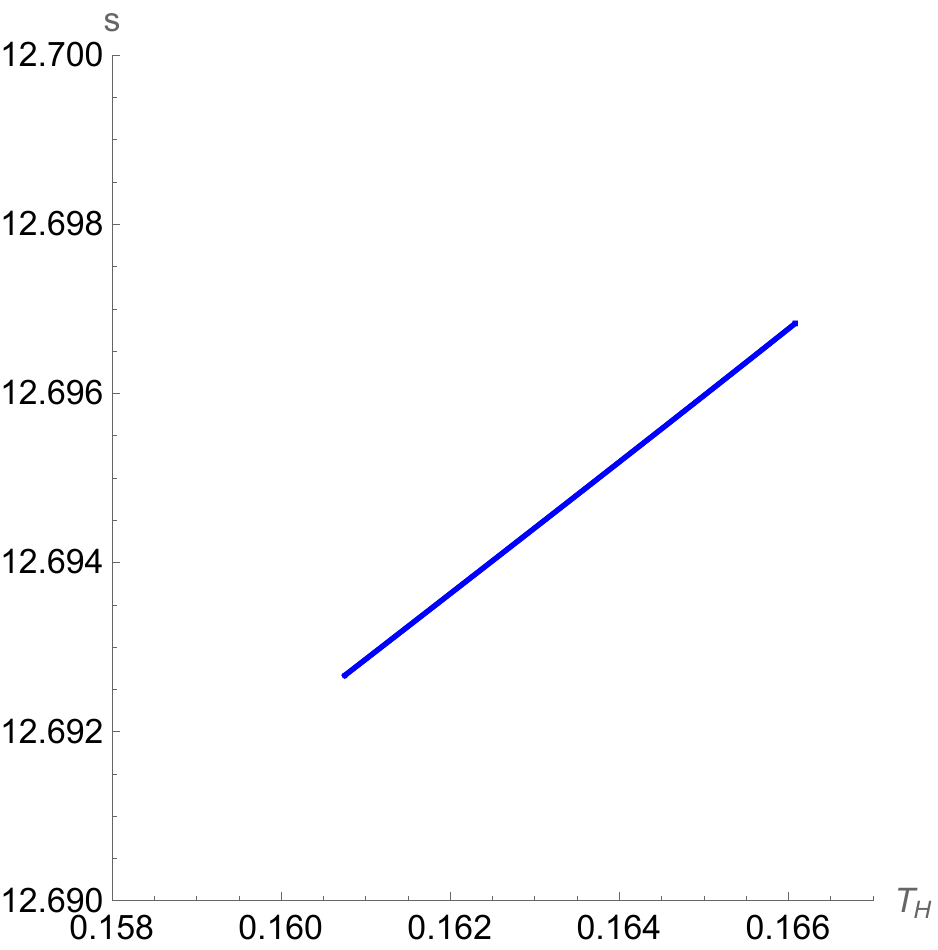}{\bf b)}
     \caption{\small Entropy density (\ref{entbh}) vs Hawking temperature (\ref{bhtemp}) for a) $a^2 =0.25$, b) $a^2=0.8$; for all we set $w_{h}=0.01$ .}
    \label{Fig: bhentropy}
\end{figure}

We show the behaviour of the entropy $s$  (\ref{entbh})  as a function of $T_H$ (\ref{bhtemp}) in Fig.~\ref{Fig: bhentropy}.

%The red curve in Figs.~\ref{Fig: %bhentropy3}-\ref{Fig: bhentropy4} %correspond to the solutions with  %$\phi_h\in(\phi_2,\phi_3)$ for  $a^2=0.8$, %the blue one is the solution with $\phi_h %\in (\phi_{i},\phi_{\pm})$, the cyan curve %describes the behaviour of the entropy on  %the  temperature for %$\phi\in(\phi_m,\phi_p)$ calculated on the %exact formulae.

\begin{figure}[h!]
    \centering
     \includegraphics[width=6.5cm]{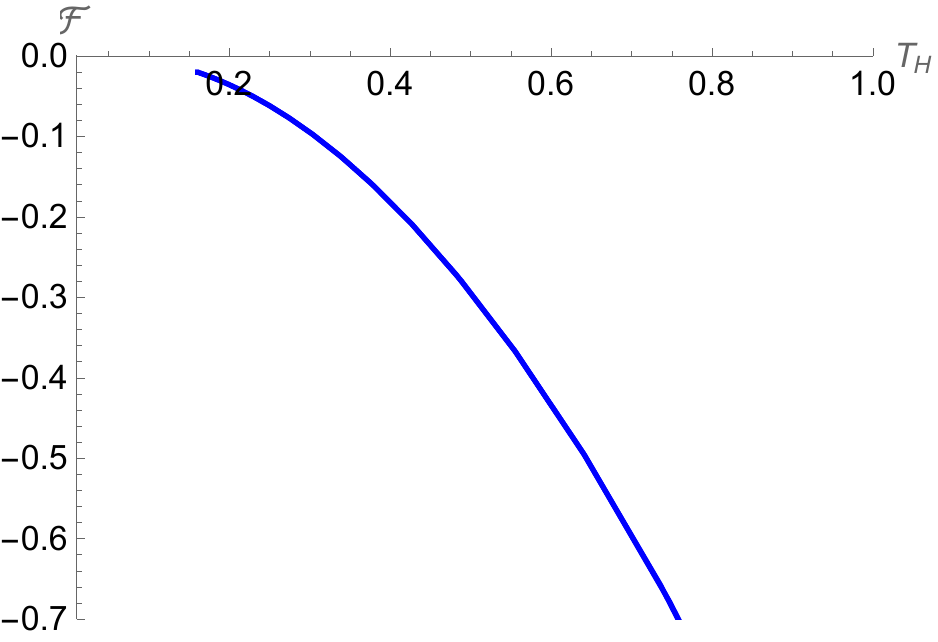}{\bf a)}
     \includegraphics[width=8.5cm]{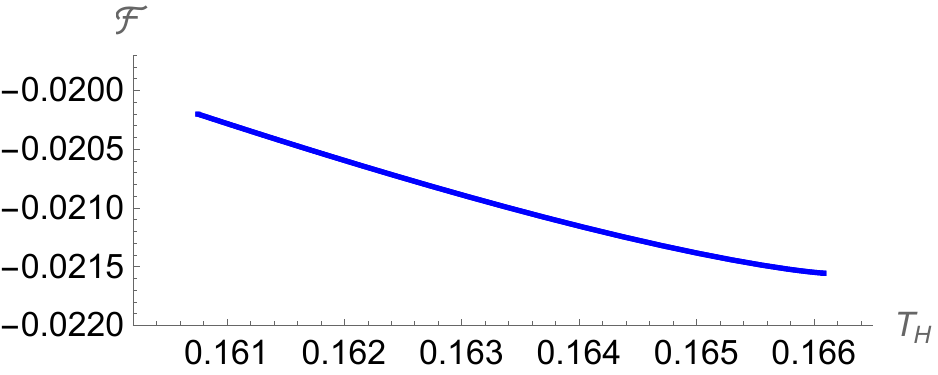}{\bf b)}
     \caption{\small Free energy (\ref{freeen}) as a function of Hawking temperature (\ref{bhtemp})  for a) $a^2 =0.25$, b) $a^2=0.8$; for all we set $w_{h}=0.01$ }
    \label{Fig: freeenTh}
\end{figure}

It is useful to look on the product of the temperature and the entropy
\be\label{sTh}
sT_{H}=\frac{2M_{p}}{\mathrm{c}_{A}}\left(1+\frac{\kappa}{\mathrm{c}_{A}} w_h\right)^{\frac{2+\kappa}{\kappa}}= \frac{2M_{p}}{\mathrm{c}_{A}}\mathsf{B}^{\frac{2+\kappa}{\kappa}}.
\ee
Taking into account \eqref{bhtemp},\eqref{entbh} and \eqref{sTh} the free energy can  be found as follows
\be\label{freeen}
\mathcal{F}=-\int sd T_{H} =-\frac{2M_{p}}{\mathrm{c}_{A}} \int^{\mathsf{B}}_{1}\frac{\bar{\mathsf{B}}^{\frac{2+\kappa}{\kappa}}}{T_{H}}\frac{dT_{H}}{d\bar{\mathsf{B}}}d\bar{\mathsf{B}} =- \frac{2M_{p}}{\mathrm{c}_{A}}\frac{\kappa+1}{\kappa+2}\left(\mathsf{B}^{1+\frac{2}{\kappa}}-1\right).
\ee
 For $\phi_h=\phi_{1}$ the thermodynamics is conformal, that can be seen  from the  expansion  of the  free  energy with $\mathsf{\Lambda^{(h)}}\simeq0$
\be\label{freeenexp}
\mathcal{F}_{\mathsf{\Lambda}^{(h)}\to 0} =-\frac{e^{2\frac{w_{h}}{\mathrm{c}_A}}}{\mathrm{c}_A}-\frac{e^{\frac{2w_h}{\mathrm{c}_A}}(\mathrm{c}_{A}^2 -2 w_h^2+2\mathrm{c}_A w_h)}{2a^2 \mathrm{c}_{A}^3}(\mathsf{\Lambda}^{(h)})^2,
\ee
i.e. from (\ref{tempexp}) and (\ref{freeenexp})  we get as expected
\be
\mathcal{F}\sim T^{d},
\ee
where $d=2$. In Fig.~\ref{Fig: freeenTh} we show the dependence of $\mathcal{F}$ on $T_{H}$.

 \newpage
 \setcounter{equation}{0}

\section{From the near-horizon region to the boundary}\label{sec:solPhiImp}

\subsection{Analytic solutions from  the horizon to the boundary}

In this section we will discuss analytic asymptotically AdS black hole solutions,  defined for both near the horizon and near the boundary regions. These solutions can be obtained using  an additional condition
\be\label{x20}
X^2\sim 0,
\ee
the validity  of using which we observed from the behaviour of numerical holographic RG flows of the system  \eqref{dz/da}-\eqref{dy/da}. In fact, the constraint (\ref{x20}) says about the slowly changing scalar field. Below we will show that the constraint (\ref{x20}) allows to reduce the dynamical system such that the equation  for the variable $Y$ decouples and the equations for the scalar field and the scale factor are  simplified.

Thus, imposing the constraint (\ref{x20}) for eqs. \eqref{dz/da}-\eqref{dy/da} we come to the system of equations
\bea\label{adz/da}
&&\frac{dz}{dA}=z(z-1)X,\\ \label{adx/da}
&&\frac{dX}{dA}=-\left(Y+2\right)\left(X+\frac{a^2}{2}\frac{V_{\phi}}{V} \right),\\ \label{ady/da}
&&\frac{dY}{dA}=-Y\left(Y+2\right) \,.
\eea

\begin{figure}[t]
    \centering
  \includegraphics[width=11cm]{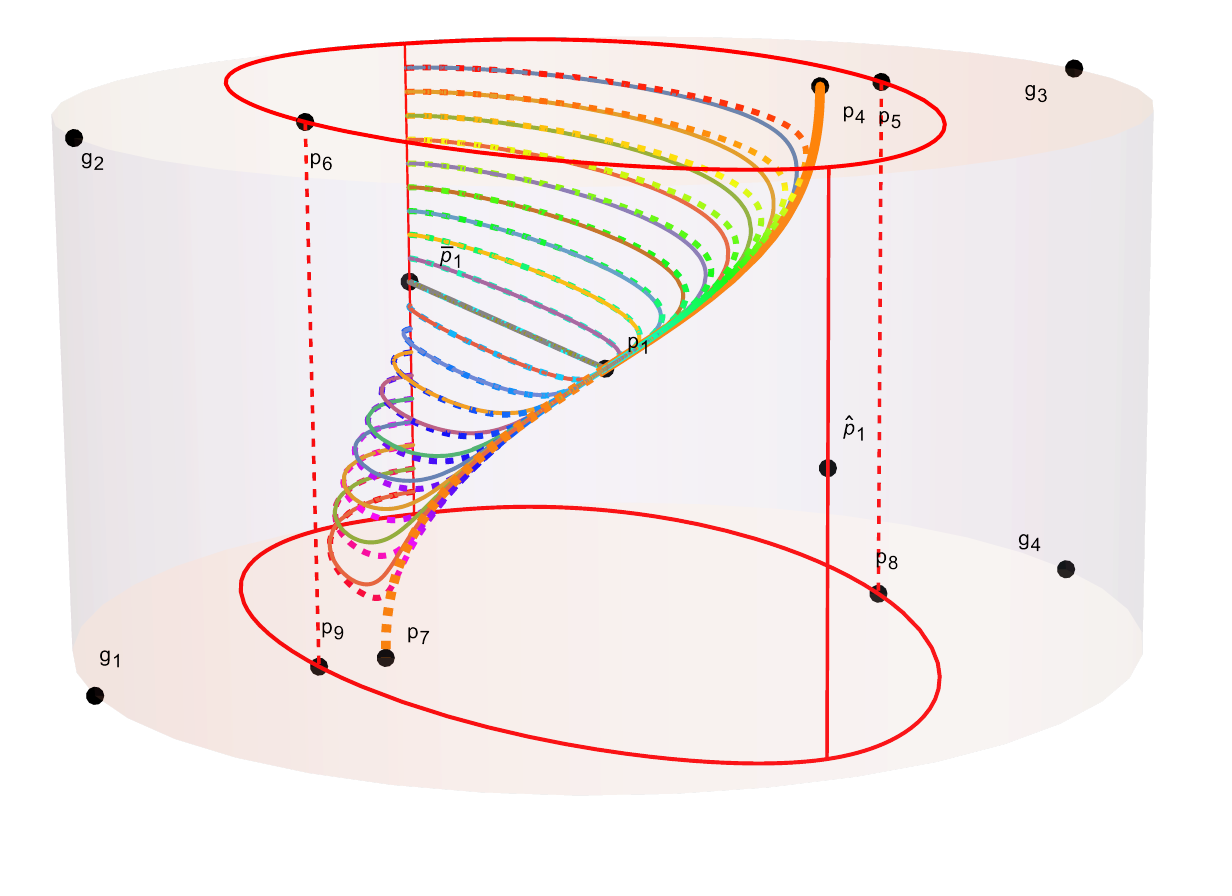}
    \caption{The numerical trajectories of the dynamical systems (\ref{dz/da})-(\ref{dy/da}) and \eqref{adz/da}-\eqref{ady/da} (dashed) in the cylinder for $a^2=0.25$.}
    \label{Fig:cylfull}
\end{figure}

Mapping the system \eqref{adz/da}-\eqref{ady/da}  into a unit cylinder, one can find that for
 the case $0<a^2<1/2$ the numerical solutions with the condition (\ref{x20}) from the near-horizon region with some $\phi_h$ to the  AdS critical point $p_{1}$ have a greater deviation from the numerical trajectories of the system (\ref{EqsonCyl001})-(\ref{EqsonCyl0013}) without additional constraints ending in $p_1$, the further the initial condition for $\phi_{h}$ from the extremum $\phi=0$ ($z=1/2$)   is chosen. In Fig. \ref{Fig:cylfull} we show the trajectories of \eqref{adz/da}-\eqref{ady/da} in the cylinder for $0<a^2\leq 1/2$. The solid curves correspond to the holographic RG flows  of (\ref{EqsonCyl001})-(\ref{EqsonCyl0013}) without additional constraints, while the dashed curves are numerical solutions with the condition (\ref{x20}).

\begin{figure}[t]
    \centering
  \includegraphics[width=11cm]{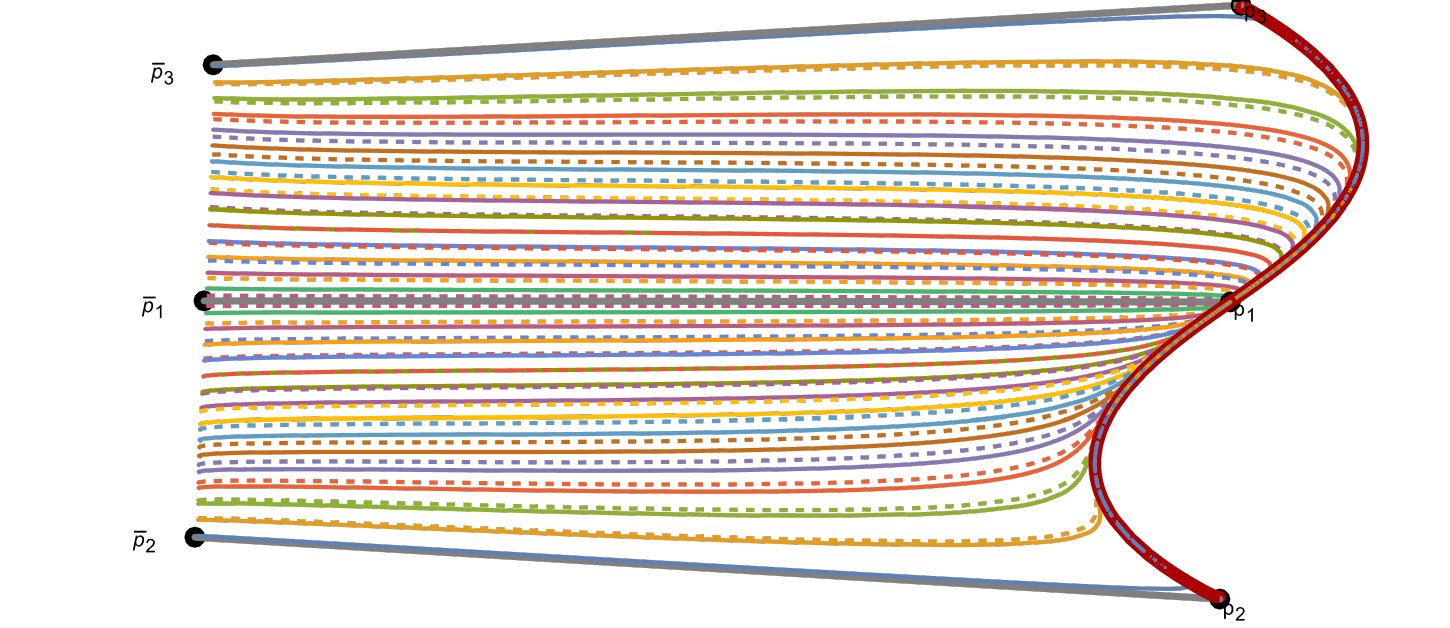}
    \caption{The numerical trajectories of the dynamical systems (\ref{dz/da})-(\ref{dy/da}) (solid) and \eqref{adz/da}-\eqref{ady/da} (dashed) for $a^2=0.8$. The values of  $\phi_h$ for the dashed flows corresponding to \eqref{adz/da}-\eqref{ady/da} slightly differ  from the initial conditions for (\ref{dz/da})-(\ref{dy/da}), otherwise they would completely merge in the figure.}
    \label{Fig:cylfullcaseTWO}
\end{figure}

To find the  analytical form of the solutions to \eqref{adz/da}-\eqref{ady/da}, we assume that the  variable $Y$ tends to infinity near the horizon, $Y(0)\to\infty$.  %\footnote{Here we take the value of the scale factor on the horizon $A_h=0$ for simplicity}.
The equation \eqref{ady/da} is independent and can be easily integrated as
\be \label{forAY}
Y(A) = \frac{2}{e^{2 (A-A_{h})}-1} \,.
\ee
Plugging \eqref{forAY} into eq.\eqref{adx/da} and returning from $z$ to the original variable $\phi$ \eqref{Zvar} the  other two equations  can be represented as a single second-order differential equation for the scalar field $\phi$ on  $A$
%\bea \label{adZ-1}
%\frac{d\phi}{dA} &=& X, \\
%\label{aDX-1}
%\frac{dX}{dA} &=& -\left(1+\coth A \right) \left(X+\frac{a^2}{2} \frac{V_{\phi}(\phi)}{V(\phi)} \right) \,,
%\eea
\be \label{a-one-eq-XF-1}
\frac{d^2 \phi}{d(A-A_{h})^2}+\left(1+\coth (A-A_{h}) \right) \left(\frac{d\phi}{d(A-A_{h})}+\frac{a^2}{2} \frac{V_{\phi}(\phi)}{V(\phi)}\right ) =0 \,.
\ee
We note that the equation \eqref{a-one-eq-XF-1}  is not expanded near the horizon, i.e. a  Taylor series  expansion of \eqref{a-one-eq-XF-1}  by $A$ near $A_h$ up to the first order yields  the  scalar field equation \eqref{one-eq-XF-1}. However, we note the expansion  of the  potential term in  \eqref{a-one-eq-XF-1} can be done if  $\phi_{h}$ is taken near $\phi_{1}$ both for $a^2<\frac{1}{2}$ and $\frac{1}{2}<a^2<1$. The scalar field corresponding to the trajectories starting from $\phi_{2,3}$ ($\bar{p}_{2,3}$), as well as for $\phi_{h}$ far from $\phi_{1}$, changes fast \cite{Golubtsova:2024dad} that precludes to use the expansion of $\frac{V_{\phi}}{V}$. Thus, we are able to construct the trajectories with the constraint (\ref{x20}) from $\bar{p}_{2,3}$ to $p_{1}$ numerically, but we cannot succeed analytically.

 Eq.\eqref{a-one-eq-XF-1} near $\phi_h$ can be brought to the hypergeometric equation
\be \label{HG}
r(1-r)\, \frac{d^2 \Phi(r)}{dr^2}+(1-2r)\,\frac{d\Phi(r)}{dr}-\frac{\mathsf{K}^{(h)}}{2} \Phi(r) =0,
\ee
where we define
\be \label{PHIde}
\Phi:=\phi-\phi_{h}+\frac{\mathsf{\Lambda}^{(h)}}{\mathsf{K}^{(h)}},
\ee
with $\mathsf{\Lambda}^{(h)}$ and $\mathsf{K}^{(h)}$ given by \eqref{SlDef} and
\be \label{th-eq-XF-t}
r=\exp{2 (A-A_{h})}.
\ee
It is worth to be  noted that we can derive eq.(\ref{HG}) from EOM for the scalar field (\ref{eqd}) under the assumption \eqref{x20}, which  it turn says that $A$ is linear.

The regular singular points $r=1$ and $r=\infty$ of eq.(\ref{HG}) correspond to the near-horizon and boundary regions, respectively.

\begin{figure}[t]
    \centering
  \includegraphics[width=6.5cm]{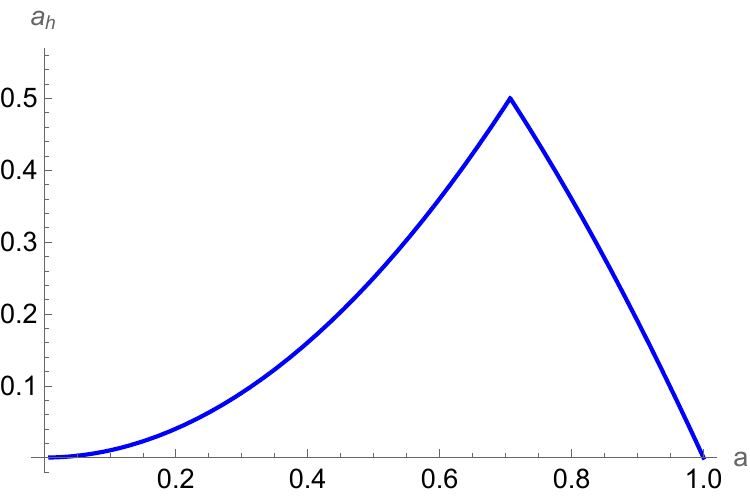}\, {\bf a)}\,\,
  \includegraphics[width=6.5cm]{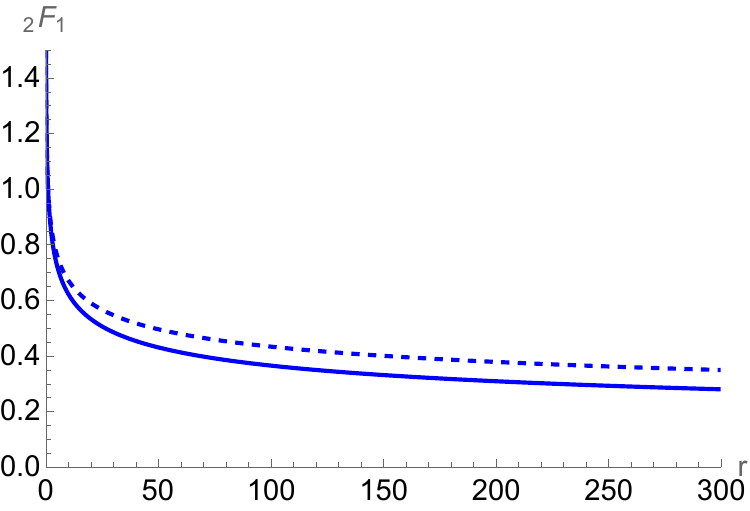}\,\, {\bf b)}
    \caption{ {\bf a)}The dependence of the parameter $a_h$ \eqref{hypgpar} on $a$; {\bf b)} The behaviour of ${}_2 F_1 (a_h,1-a_h,1,1-r)$ on $r$ for $a^2=0.25$ (solid), $a^2=0.8$ (dashed).}
    \label{Fig:hypergeomfunc}
\end{figure}

Then,  the solution for the scalar field  near the horizon can be read off immediately  from the fundamental solution to eq. (\ref{HG}) near $r=1$, namely,
\be \label{2f1}
\Phi(r)={}_2 F_1 (a_h,1-a_h,1,1-r) \,,
\ee
where  $a_{h}$ is
\be\label{hypgpar}
a_h=\frac{1}{2}\left(1-\sqrt{1-2\mathsf{K}^{(h)}}\right).
\ee
The  solution \eqref{2f1}  is divergent at $r=0$, which  is not of our interest, but finite in the region  from $r=1$ to $r=\infty$. We show the behaviour of the solution \eqref{2f1}  in Fig.~\ref{Fig:hypergeomfunc} {\bf b)} for $a^2=0.25$ (solid) and  $a^2=0.8$  (dashed).

% the solution for the scalar field (\ref{PHIde}) with (\ref{2f1}) is related with a branch of the solutions \eqref{solDilStar} given by Bessel functions of the first
%kind.

 %This equation corresponds to the system \eqref{adz/da}-\eqref{ady/da}, for which the function $\mathcal{C}_{(z,a)}$ is expanded in series by $z$ near $z_h$ up to the first order.
As for the metric function  $A$, the
 constraint (\ref{x20}) leads to the fact that the equations of motion \eqref{eom1}-\eqref{eom3}  are simplified and we have  for $A$
\be\label{eqAsolve}
 \ddot{A} =0.
\ee
Then, the solution to (\ref{eqAsolve}) is given by
\be\label{scalefA}
A = \mathfrak{c}_{A}w + \mathrm{c}_{2},
\ee
where the constants of integration  are chosen as
\be\label{cAfrak}
\mathfrak{c}_{A} = \sqrt{-\frac{V(\phi_h)}{2}},\quad \mathrm{c}_{2}=0.
\ee

The blackness function $f$ can be found from (\ref{forAY}) and the definition of $Y$ \eqref{XvarDef}

\be
f=e^{\mathrm{c}_{g}}\left(1-e^{-2\mathfrak{c}_{A}(w -w_h)}\right),
\ee
so the constant $\mathrm{c}_{g}$ should be fixed from the equations of motion  \eqref{eom1}-\eqref{eom3} as  $\mathrm{c}_{g} =0$.

Taking into account (\ref{scalefA})-(\ref{cAfrak})  the solution for thee scalar field takes the form
\be
\phi = \phi_{h}+{}_2 F_1\left(
\frac{1-\sqrt{1-2 \Delta^{(h)}}}{2},\frac{1+\sqrt{1-2 \Delta^{(h)}}}{2},1,1-e^{\mathfrak{c}_{A}(w-w_h)}\right).
\ee

It is interesting to note that the constraint \eqref{x20} brings us that the metric of the solution matches with the metric of the non-rotating BTZ black hole (\ref{BTZbh})-(\ref{BTZbh2}), while the scalar field is a solution to the hypergeometric equation (\ref{HG}).  This equation appears if one considers merely a scalar field in the BTZ black hole background. So this class of the solutions has a conformal thermodynamics. It worth also to be noted that in \cite{Arkhipova:2024iem}
it was shown that constraint \eqref{x20} is related with the condition for vanishing of  the $c$-function.

\subsection{The behaviour of the scalar field near boundary}

Despite the solution for the scalar field (\ref{2f1}) is obtained for the region near the horizon, from Fig.~\ref{Fig:cylfull} we can observe that such flows  end in the AdS critical point $p_1$,  if the value of the scalar field on the  horizon in taken in the neighbourhood of the extremum $\phi=0$. This happens because  the hypergeometric function (\ref{2f1}) can be extended from the region $r=1$ to $r\to+\infty$.

Consider the behavior of the scalar field  (\ref{PHIde})  with (\ref{2f1}) near the boundary  $r\to \infty$ , taking into account $\mathsf{\Lambda}^{(h)}\sim 0$.
The Taylor series expansion of (\ref{2f1}) near the boundary $r\to +\infty$ leads to
\be\label{PhiAinf-a}
\phi_{A\to\infty}\simeq\frac{\Gamma (\Delta_{-}-1)}{\Gamma(\frac{\Delta_{-}}{2})^2}e^{- \Delta_{+}w/\ell}+\ldots+
\frac{\Gamma (\Delta_{+}-1)}{\Gamma (\frac{\Delta_{+}}{2})^2}e^{-\Delta_{-}w/\ell}+\ldots,
\ee
where we used that near $\phi_{h}=0$ the integration constant $\mathfrak{c}_{A}$ (\ref{cAfrak}) can be related with the radius of the $AdS$ spacetime in $\phi=0$
\be
\mathfrak{c}_{A} =1/\ell.
\ee
The conformal dimensions $\Delta_{\pm}$ of the dual operator are defined by
\be
\Delta_{\pm}=1\pm\sqrt{1+M^2\ell^2}=1\pm\sqrt{1+\frac{a^2}{2}V_{\phi\phi}(\phi_{1})\ell^2},
\ee
with  $M^2 = V_{\phi\phi}(\phi_{1})$,
i.e. we used
\be
1\pm\sqrt{1-2 \Delta^{(h)} }%&=&1\pm\sqrt{1-2\frac{a^2}{2}\frac{V_{\phi\phi}(\phi_{h})}{V({\phi_{h})}}}\\
%=1\pm\sqrt{1+\frac{a^2}{2}\frac{V_{\phi\phi}(\phi_{h})}{\mathfrak{c}_{A}^2}}
= 1\pm\sqrt{1+M^2_{\phi}\ell^2}.
\ee
%and
%\be
%M^2_{\phi} = 8a^2(a^2-1)
%\ee
Moreover, the coefficients in  (\ref{PhiAinf-a}) coincide with those from \cite{Balasubramanian:1998sn}.
The form (\ref{PhiAinf-a}) matches with  the statement that the scalar field near the boundary should has  the following behaviour
\be
\phi\simeq \phi_{-}e^{-\Delta_{-}w}+\ldots+\phi_{+}e^{\Delta_{+}w/\ell}+\ldots.
\ee
It is interesting to note that the BTZ black hole in the Poincare like coordinates  can be related with the AdS$_{3}$ spacetime. Thus, coming to the Poincar\'e coordinates in (\ref{2f1}) and doing the change $r^2_{h}=-1$ we get an answer for the scalar field in the AdS$_3$ spacetime:
\be
\Phi(r)={}_2 F_1 \left(\frac{\Delta_{-}}{2},\frac{\Delta_{+}}{2},1,1+r^2\right) \,,
\ee
that is in agreement with an analysis from \cite{Freedman:1998tz,Balasubramanian:1998sn} up to  change of the radial variables (as $\tan^{-1} r = \rho$, $\tan\rho=\sinh\mu$).

\section{Discussion}\label{sec:discussion}

In this  work, we have studied holographic renormalization group flows at finite temperature in the 3-dimensional truncated supergravity with a scalar potential. These RG flows  are thermal generalizations of RG flows from \cite{Arkhipova:2024iem} triggered by either  relevant deformations or non-zero VEV of the operator.

For the  potential with one extremum ($a^2\leq\frac{1}{2}$), the constructed thermal holographic RG flows correspond to the deformation of the dual theory by the relevant operator. The flows are described by asymptotically AdS black holes with with the scalar field, which behaves monotonically and $\phi_h$ at the horizon can take any value from the domain of definition of the potential. For the case of the  potential with several extrema ($\frac{1}{2}\leq a^2<1$) , we have constructed  numerically thermal holographic RG flows, which are related with deformation of the dual conformal theory both by the relevant operator and by a nonzero VEV of the scalar operator. The former flows are characterized by a monotonic scalar field, as we have seen for the case of one extremum, $a^2\leq\frac{1}{2}$. For them  the value of the scalar field at the horizon $\phi_h$ can take a value between the extrema of the potential.
RG flows associated with a fixed point deformation by a non-zero expected mean of the operator are characterized by a non-monotonic behavior of the scalar field associated with a non-renormalizable contribution on the conformal boundary. It is shown that the scalar field on the horizon for such flows takes values near additional extrema.

For the thermal RG flows related with the relevant deformations we have found analytic solutions  in the near-horizon regions. Using them we have studied the thermodynamics and have shown that at the extrema of  the potential the thermodynamics is conformal.

For the monotonic slowly changing scalar field we have found a special class of the flows, which geometry is described by the non-rotating BTZ metric, and the solution for the scalar field is a solution of the hypergeometric equation. From the numerical observations we have shown that the solution for a scalar field can be extended from the horizon to the boundary. From the expansion of the scalar field solution on the boundary we have seen the correct behaviour.
%This class of RG flows is associated with the condition on the vanishing of the Zamolodchikov $c$-function.

It will further be interesting, to explore thermal two-point correlation functions \cite{Rodriguez-Gomez:2021pfh,Rodriguez-Gomez:2021mkk} in these geometries.

\section*{Acknowledgments}
AG and MP would like to thank Irina Ya. Aref'eva and Lev Astrakhantsev for useful discussions. The work of AG is supported by the Russian Science Foundation under Grant Number  20-12-00200.

\newpage

\appendix

\setcounter{equation}{0} \renewcommand{\theequation}{A.\arabic{equation}}

\section{Exact holographic RG flow}\label{app:AppA}
\setcounter{equation}{0}\renewcommand{\theequation}{A.\arabic{equation}}
For the model (\ref{act}) the exact domain wall solution, which preserve half of supersymmetries exists and was obtained in the work \cite{Deger:2002hv}.
The scale factor of this solution reads
\be\label{degerSF}
A_{\rm susy} = \frac{1}{4a^2}\ln(e^{4a^2w}-1),
\ee
the scalar field is given by
\be\label{degerScF}
\phi_{\rm susy} =  \frac{1}{2}\ln\left(\frac{1+e^{-2a^2w}}{1-e^{-2a^2w}}\right),
\ee
where the radial coordinate $w$ runs from $0$ to $\infty$.\\

The metric of the domain wall solution with (\ref{degerSF}) can be represented as follows
\be\label{metricDeg}
ds^2_{\rm susy} = \left(e^{4a^2w} -1 \right)^{\frac{1}{2a^2}}\left(-dt^2+dx^2\right) +dw^2.
\ee
The domain wall solution (\ref{metricDeg}) with (\ref{degerScF}) has an AdS asymptotics with $w\to \infty$, while if one send  $w$ to zero, the  solution  has a singular geometry. For $a^2\leq\frac{1}{2}$  this singular behaviour turns to be acceptable since the solution satisfies the Gubser's bound.

\be
\phi\simeq \phi_{UV} + j\ell^{d-\Delta}e^{-(d-\Delta)w/\ell},\quad w\to \infty,
\ee
where in our case an UV fixed point is at  $\phi=0$, $\Delta = 1\pm |1-2a^2|$, so if $a^2<\frac{1}{2}$ the conformal dimension $\Delta = 2(1-a^2)$ and if  $\frac{1}{2}<a^2<1$ we have $\Delta = 2a^2$.
\bea
\Delta_{-}=d-\Delta &=& 2a^2, \quad \textrm{if}\quad a^2<\frac{1}{2}, \\
\Delta_{-}=d-\Delta &=&2-2a^2,\quad  \textrm{if}\quad \frac{1}{2}<a^2<1.
\eea
The  parameter $\ell$ is the radius of the AdS spacetime
$\ell = \sqrt{-\frac{2}{V(\phi_{UV})}}$.
Then one can write
\be
\phi\simeq \phi_{UV} + j
\ell^{2a^2}e^{-2a^2w/\ell},\quad w\to \infty,
\ee
In  our case $V(\phi_{UV})=V(0)=2\Lambda$. Setting $\ell_{UV}=1$ , we get $\Lambda=-1$.  The Taylor series expansion of the Deger solution gives
\be
\phi\simeq \phi_{UV} +
e^{-2a^2w},\quad w\to \infty.
\ee

\setcounter{equation}{0} \renewcommand{\theequation}{B.\arabic{equation}}

\section{BTZ black hole in various coordinates}

The metric of the non-rotating BTZ black hole in the dowain-wall coordinates
\be
ds^2 = e^{2\sqrt{-\frac{V(\phi_{h})}{2}} \, w \,} \left(-(1-e^{-\sqrt{-2 \, V(\phi_{h})} (w-w_h)})dt^2 + d\phi^2 \right) +\frac{dw^2}{1-e^{-\sqrt{-2 \, V(\phi_{h})} (w-w_h)}}.
\ee
where  the blackenning function  given by
\be
\quad f(w) = 1-e^{-\sqrt{-2 \, V(\phi_{h})} (w-w_h)}.
\ee
with $\ell =1$.
Coming to the Poincar\'e coordinates with
$dw =\frac{dr}{r}$,
we get
\be
ds^2 = e^{2\ln r} \left(-(1-e^{-2 (\ln r-\ln r_h)})dt^2 + d\phi^2 \right) +\frac{dr^2}{r^2(1-e^{-2 (\ln r-\ln r_h)})}.
\ee
that can be brought into the following form
\be
ds^2 =  -(r^2- r^2_h) dt^2 + r^2dx^2  +\frac{dr^2}{(r^2- r_h^2)}.
\ee
Remembering that the AdS$_{3}$ spacetime in terms of global coordinates with $\ell=1$ has  the form
\be\label{ads3glob}
ds^2 = -(r^2+1)dt^2  +r^2d\varphi^2 + \frac{dr^2}{r^2+ 1}.
\ee
%Thus, we should set
%\be
%r_{h}^2 = -1
%\ee
%to get (\ref{ads3glob}).

%\be
%\Phi(r)={}_2 F_1 (a_h,1-a_h,1,1-e^{2\mathfrak{c}_{A}(w-w_h)}) \,,
%\ee
%where  $a_{h}$ is
%\be
%a_h=\frac{1}{2}\left(1-\sqrt{1-2\mathsf{K}^{(h)}}\right),\quad  \mathsf{K}^{(h)}=\frac{a^2}{2}\frac{V_{\phi\phi}(\phi_{h})}{V(\phi_{h})}.
%\ee

%\be
%\Phi(r)={}_2 F_1 \left(a_h,1-a_h,1,1-\left(\frac{r}{r_h}\right)^2\right) \,,
%\ee

%\be
%\Phi(r)={}_2 F_1 \left(a_h,1-a_h,1,1+r^2\right) \,,
%\ee

\newpage
\bibliography{bib.bib}
\bibliographystyle{utphys.bst}

\end{document}